\documentclass[conference, a4paper]{IEEEtran}
\usepackage[left=1.32cm,right=1.32cm,top=1.9cm,bottom=4.3cm]{geometry}
\usepackage{subcaption}
\usepackage{cite}
\usepackage{amsmath,amssymb,amsfonts}
\usepackage{graphicx}
\usepackage{textcomp}
\usepackage{tikz}
\usepackage{adjustbox}
\usepackage{multirow}
\usepackage{multicol}
\usepackage[affil-it]{authblk} 

\usepackage{xcolor}
\usepackage{pgfplots}
\pgfplotsset{compat=newest,compat/show suggested version=false}
\usepackage{booktabs}
\usepackage{algpseudocode}
\usepackage[ruled, lined, longend, linesnumbered]{algorithm2e}
\usepackage[utf8]{inputenc}
\usepackage{tikz}
 
\usepackage{amssymb}
\usepackage{pifont}

\def\BibTeX{{\rm B\kern-.05em{\sc i\kern-.025em b}\kern-.08em
    T\kern-.1667em\lower.7ex\hbox{E}\kern-.125emX}}
\begin{document}




\title{BARTPredict: Empowering IoT Security with LLM-Driven Cyber Threat Prediction}





\author[1]{Alaeddine Diaf}
\author[1,3]{Abdelaziz Amara Korba}
\author[2,4]{Nour Elislem Karabadji}
\author[3]{Yacine Ghamri-Doudane}

\affil[1]{Laboratoire Reseaux et Systemes (LRS), Faculty of Technology, Badji Mokhtar-Annaba University , Algeria}
\affil[2]{National Higher School of Technology and Engineering -Annaba, Algeria}
\affil[4]{Laboratoire De Technologies Des Systemes Energetiques (LTSE), E3360100, Annaba, Algeria.}
\affil[3]{L3I, University of La Rochelle, France}

\maketitle
\begin{abstract}
The integration of Internet of Things (IoT) technology in various domains has led to operational advancements, but it has also introduced new vulnerabilities to cybersecurity threats, as evidenced by recent widespread cyberattacks on IoT devices. Intrusion detection systems are often reactive, triggered by specific patterns or anomalies observed within the network. To address this challenge, this work proposes a proactive approach to anticipate and preemptively mitigate malicious activities, aiming to prevent potential damage before it occurs. This paper proposes an innovative intrusion prediction framework empowered by Pre-trained Large Language Models (LLMs). The framework incorporates two LLMs: a fine-tuned Bidirectional and Auto-Regressive Transformers (BART) model for predicting network traffic and a fine-tuned Bidirectional Encoder Representations from Transformers (BERT) model for evaluating the predicted traffic. By harnessing the bidirectional capabilities of  BART the framework then identifies malicious packets among these predictions. Evaluated using the CICIoT2023 IoT attack dataset, our framework showcases a notable enhancement in predictive performance, attaining an impressive 98\% overall accuracy, providing a powerful response to the cybersecurity challenges that confront IoT networks.

\end{abstract}



\section{Introduction}
The Internet of Things (IoT) technology has become widespread, enabling a connected network of devices that enhance operational efficiency and productivity in various aspects of daily life. However, this surge in adoption has also led to an increase in cybersecurity threats, as malicious actors target vulnerable IoT devices with sophisticated attacks, exploiting their poor security configurations \cite{10621562}. One notable example is the Mirai botnet attack in 2016, which exploited compromised IoT devices to orchestrate large-scale distributed denial-of-service (DDoS) attacks \cite{10592525}, causing widespread disruptions and resulting in estimated damages totaling hundreds of millions of dollars. 

Intrusion Detection Systems (IDSs) play a crucial role in identifying potential security risks within IoT environments. In recent years, there has been a growing interest in the development of AI-powered IDSs \cite{labiod2022fog,10496859, 10592500}, driven by their capacity to detect diverse and sophisticated attack patterns with high accuracy. Recent advancements in AI-based IDSs offer the promise of real-time intrusion detection capabilities, providing organizations with the ability to proactively identify and mitigate security threats \cite{ozkan2024comprehensive}. However, the reactive approach to threat detection employed by existing IDSs \cite{10279368, 10575541} is inadequate for addressing the dynamic and evolving nature of IoT cyber threats. Focusing solely on post-facto detection often results in significant damage occurring before suitable mitigation measures can be implemented \cite{kaur2023artificial}. 
Transitioning to proactive security strategies is crucial in cybersecurity to anticipate and counter intrusions, preempt potential threats, and reduce the likelihood and severity of successful attacks. This approach fortifies organizational security measures, enhancing overall cybersecurity readiness against evolving threats. 

In recent years, Pre-trained Large Language Models (LLMs) have emerged as a groundbreaking approach in cybersecurity. A growing body of research \cite{nam2021intrusion, omar2023vuldetect, aghaei2022securebert, vubangsi2024bert, liu2024semalbert} has extensively investigated the potential of LLMs for a wide range of threat detection applications. Harnessing their transformer-based architecture capabilities, LLM models like Bidirectional Encoder Representations from Transformers (BERT) \cite{kenton2019bert} and Bidirectional and Auto-Regressive Transformers (BART) \cite{lewis2019bart} dissect network behaviors and deviations, thus facilitating the development of proactive threat detection strategies. In this paper, we introduce an innovative intrusion prediction framework focusing on network packets, empowered by two cutting-edge LLMs. Our framework capitalizes on the autoregressive capabilities of LLMs to predict next network packets based on current ones, while also harnessing their bidirectional capabilities for packet classification. Additionally, we utilize the BERT’s purely bidirectional contextual understanding for assessing the predicted network packets. This framework provides a comprehensive understanding of network packet characteristics, ensuring precise prediction of their subsequent packets, and effectively anticipating intrusions. To the best of our knowledge, this represents the first attempt to utilize LLMs for network intrusion prediction. We fine-tune the BART model on both benign and malicious network traffic to predict next network packets, and further fine-tune it to specifically target intrusion prediction. Additionally, we fine-tune the BERT model for a packet-pair classification task to assess the predicted packets. Our experimental evaluation demonstrates the effectiveness of our proposed framework, achieving an overall accuracy of 98\% in detecting intrusions.

The remainder of this paper is structured as follows: Section II provides an overview of related work. Section III explores the methodology of the proposed framework. Section IV outlines the results of the performance evaluation, and finally, Section V concludes the paper.


\section{Related Work} \label{RT}

LLMs have emerged as powerful tools in addressing cybersecurity challenges, offering novel approaches for intrusion detection, vulnerability identification, and anomaly detection. Several studies have explored the integration of LLMs with deep learning techniques to enhance intrusion detection systems (IDSs). For instance, Nam et al. \cite{nam2021intrusion} leveraged bidirectional GPT networks to consider both past and future contexts for improved attack detection accuracy. However, the bidirectional fusion of GPT networks may escalate computational complexity and extend detection time, posing challenges for real-time applications. In the realm of vulnerability detection, Omar et al. \cite{omar2023vuldetect} proposed VulDetect, a vulnerability detection framework based on fine-tuned GPT models. With an impressive accuracy rate of up to 92.65\%, VulDetect effectively identifies software vulnerabilities. Similarly, Aghaei et al. \cite{aghaei2022securebert} introduced SecureBERT, a domain-specific LLM tailored for cybersecurity tasks. By utilizing customized tokenizers and altered pre-trained weights, SecureBERT outperforms existing models in capturing text connotations in cybersecurity text. LLMs have also shown promising results in identifying abnormal network behaviors. For instance, Vubangsi et al. \cite{vubangsi2024bert} proposed BERT-IDS, a model that analyzes network traffic and detects anomalies based on BERT embeddings. Despite its effectiveness in capturing complex patterns, BERT-IDS may require large amounts of training data for optimal performance.
Moreover, Liu et al. \cite{liu2024semalbert} introduced SeMalBERT, a fine-tuned BERT model for identifying harmful software in Windows environments. By recognizing behavioral patterns of API call sequences, SeMalBERT enhances malware detection capabilities.

While current LLM-based cybersecurity solutions excel in detecting known attacks and stopping ongoing threats, they often overlook predicting multistage attacks and preventing catastrophic damages. Therefore, there is a need for innovative approaches that address these limitations and enhance proactive threat prediction capabilities. To bridge this gap, we propose a novel network packet-based intrusion prediction framework designed using BART and BERT LLMs. Our framework aims to predict intrusions, including multistage and unseen attacks, based on current network packets, thereby advancing proactive threat detection in cybersecurity.

\section{Proposed Solution}\label{SOL}

This section outlines the essential steps of our proposed framework, which consists of two core components: packet parsing and pre-processing, and fine-tuning pre-trained LLMs to predict and classify next packets.

\begin{figure*}
    \centering
    \includegraphics[scale=0.32]{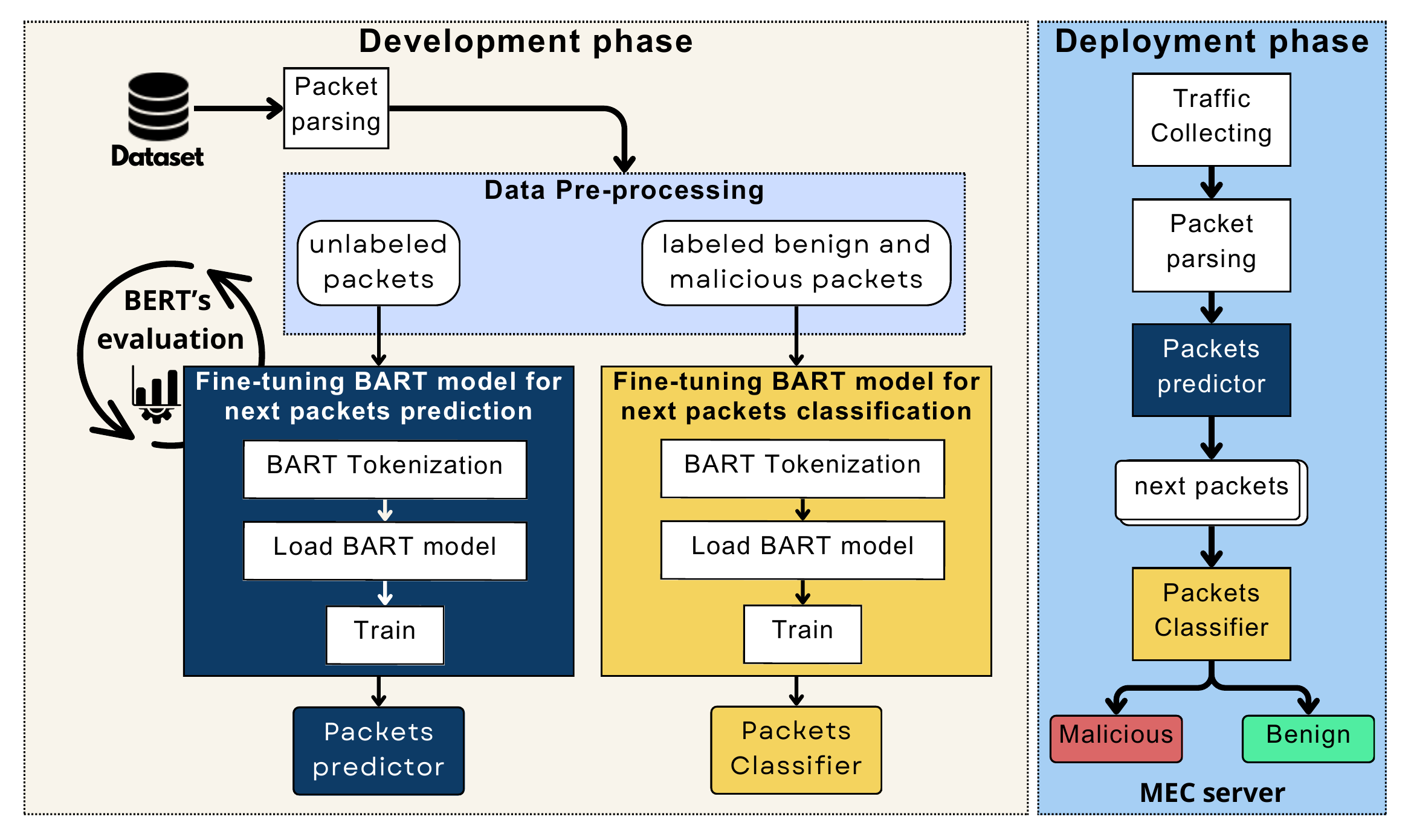}
    \caption{Workflow of the proposed intrusion prediction framework }
    \label{fig:archi3}
\end{figure*}

During the development of our intrusion prediction framework (refer to Figure \ref{fig:archi3}), our focus is on fine-tuning two separate BART models for predicting and classifying next packets, while also fine-tuning BERT to evaluate prediction accuracy. In IoT networks, packets form a sequential data flow, where each packet's features evolve over time, influenced by both preceding and subsequent packets. This sequential representation of network traffic is pivotal for predicting and detecting potential intrusions.

Initially, we harness the autoregressive capabilities of BART to capture historical context and address long-range dependencies when predicting next packets. Moreover, as BART seamlessly integrates both autoregressive and bidirectional capabilities, we fine-tune it for packet classification, exploiting its ability to learn prolonged dependencies. This aids in identifying normal and malicious traffic patterns, thereby enhancing network security measures. Evaluating BART's predictions for next packets necessitates a comprehensive understanding of the context surrounding both packets to accurately infer their relationship. To achieve this, we harness BERT's bidirectional contextual comprehension by fine-tuning it for a packet-pair classification task aimed at determining whether the predicted packet follows a given packet, considering BERT's proven effectiveness in this regard.




During the deployment phase (see Figure \ref{fig:archi3}), after packet collection and parsing, the BART-based next packets predictor predicts next packets, which are then classified by the BART-based packets classifier as either normal or malicious. This deployment is strategically situated at the Multi-Access Edge Computing (MEC) server level, enhancing overall effectiveness and responsiveness. By processing packets at the MEC server level, we achieve low-latency processing and reduce network congestion, optimizing security measures for IoT networks. This approach also relieves IoT devices of the burden of hosting security frameworks, as they often have limited computing capabilities and resources.

\subsection{Packet Parsing and Pre-processing}
The proposed framework predicts network intrusions by analyzing packet headers, extracting relevant features from captured packets, and parsing various application protocols. By focusing on packet headers instead of the entire payload, the framework streamlines processing, addresses privacy concerns, and mitigates privacy issues associated with IoT devices. This approach ensures intrusion prediction effectiveness while safeguarding user privacy. After packet parsing and pre-processing, the framework provides suitable input for LLM models.\par
\subsection{Transformer-Based Intrusion Prediction} 

Network traffic consists of a series of packets, each conveying data between network nodes. The integration of LLMs in the framework, particularly sequence-to-sequence models based on the transformer architecture \cite{vaswani2017attention}, is motivated by the complex nature of network traffic analysis. The transformer architecture, utilizing the self-attention mechanism, enables parallel processing of sequential data, extracting significant patterns. This design makes transformer-based models ideal for next packets prediction, assessing and classification. Given an input vector representing a network packet, the mathematical expression for the self-attention mechanism is: \begin{equation} Attention(Q, K, V ) = softmax \left( \frac{QK^{T}}{\sqrt{d_{k}}} \right )V \end{equation} where Q, K, and V are the query, key, and value matrices, respectively, and $d_k$ represents the dimensionality of the key vectors. The attention mechanism enables the creation of context-aware representations that capture long-range dependencies efficiently. In developing a robust next packet predictor using LLMs, we implement a feedback loop that integrates two fine-tuned models: BART for predicting network traffic and BERT for evaluating the predicted traffic. Additionally, a distinct BART model is fine-tuned for packet classification, enhancing the framework with a specialized packet classifier.\par

\subsubsection{LLMs-based next packets prediction}

Initially, we fine-tune the BART model with a network packets dataset to produce accurate next packets based on current packets, ensuring contextual consistency. BART aims to optimize the probability of the next token in a sequence by leveraging its autoregressive decoder and considering the context established by preceding tokens as follows:
\begin{equation} \mathcal L_{BART decoder}(\theta)= arg max_\theta\sum_{i=1}^N\operatorname{log}P{(ti\vert t_{<i};\operatorname\theta)}\end{equation}
By adjusting the model parameters $\theta$ across the entire training dataset that comprises $N$ tokens. Where $t_{<i}$ represents the context of tokens before $t_i$ that represents the $i-th$ token.

\begin{figure}
    \centering
    \begin{subfigure}[b]{0.32\textwidth}
        \includegraphics[width=\textwidth]{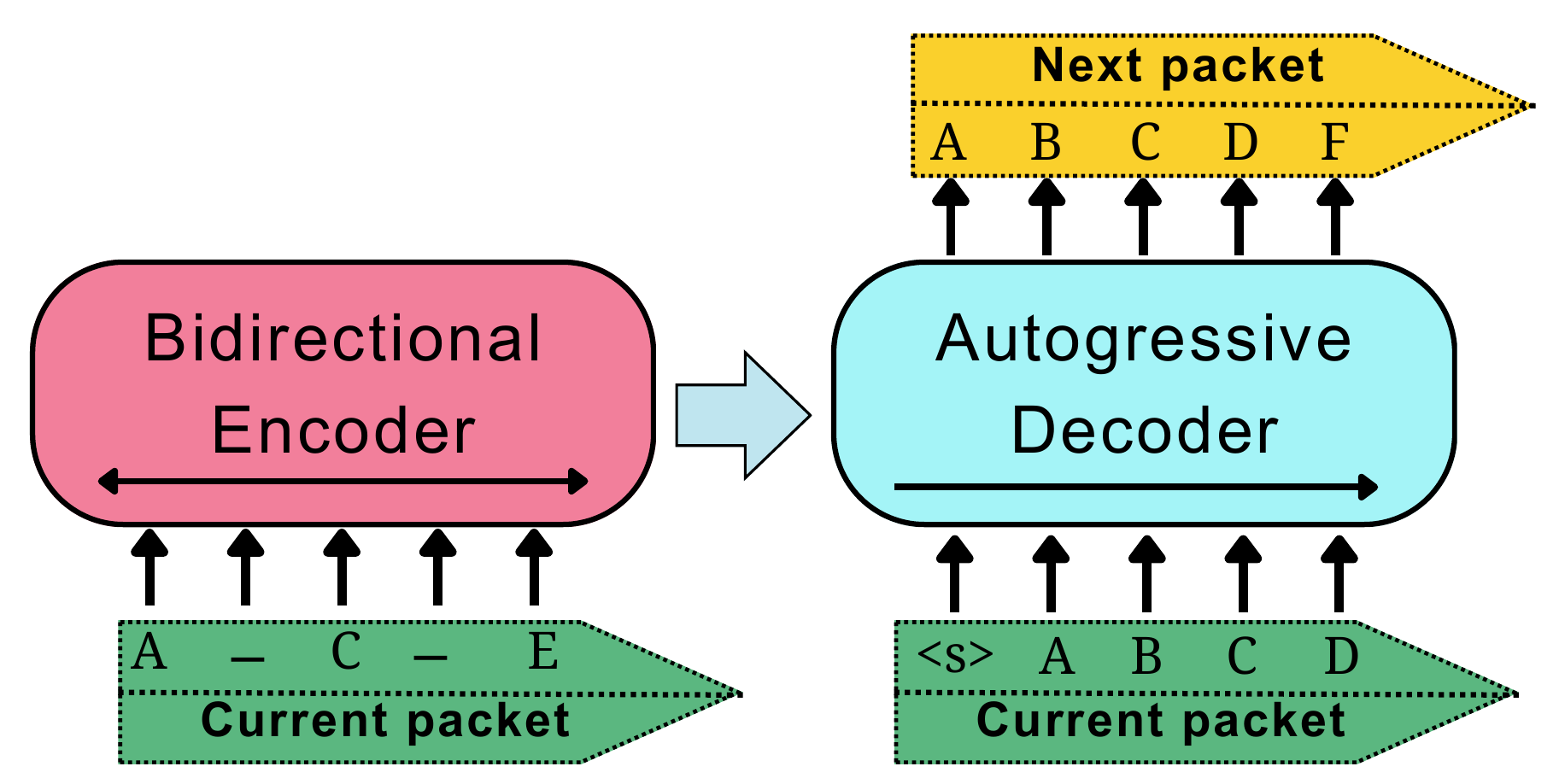}
        \caption{Modified architecture  of BART model with network packets as input and output for next packets prediction task (taken from \cite{abdulganiyu2023systematic}).}
        \label{fig:BARTarchi_a}
    \end{subfigure}
    \hfill
    \begin{subfigure}[b]{0.32\textwidth}
        \includegraphics[width=\textwidth]{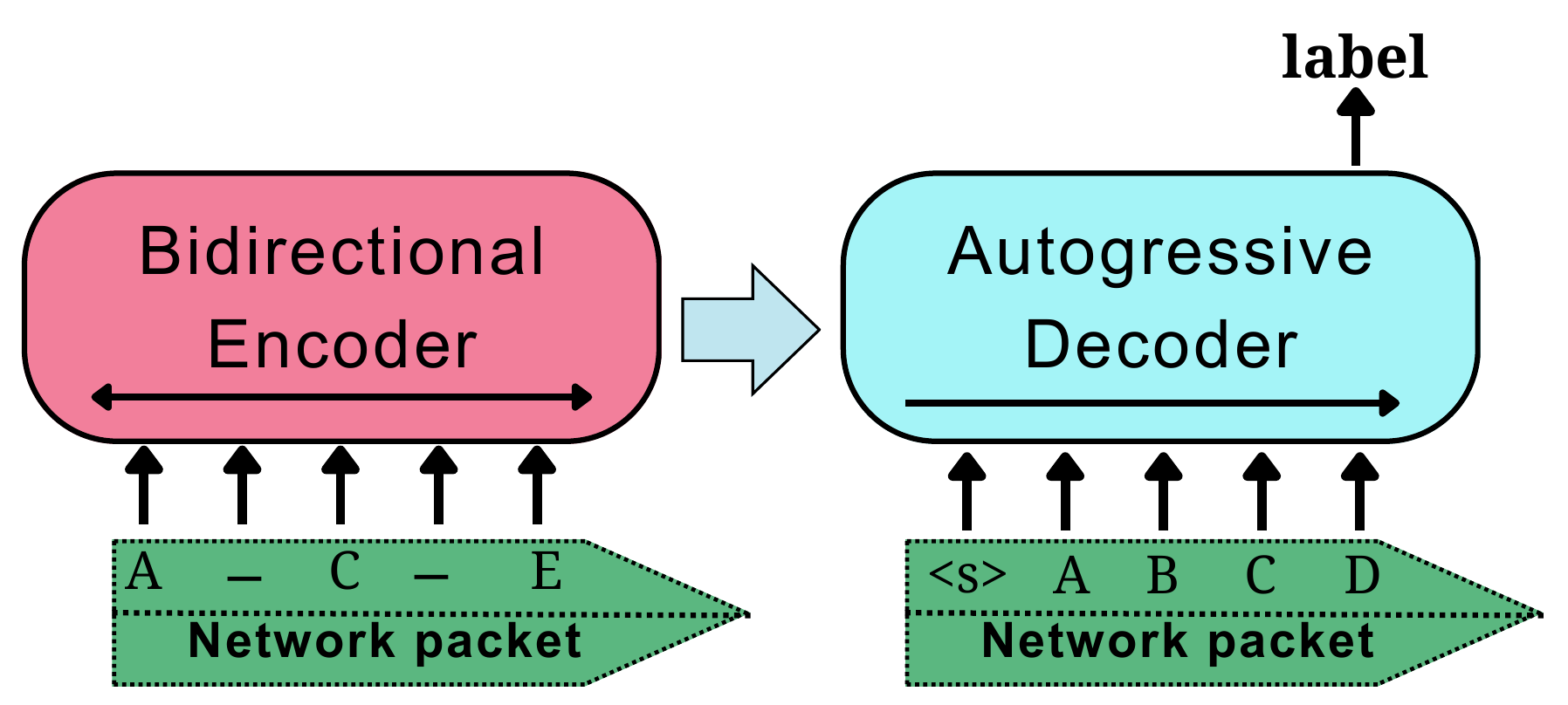}
        \caption{Modified architecture of BART model with network packets as input for packets classification task (taken from \cite{abdulganiyu2023systematic}).}
        \label{fig:BARTarchi_b}
    \end{subfigure}
    \caption{Modified structure of BART with network packets as input for Intrusion Prediction task (taken from \cite{abdulganiyu2023systematic}).}
    \label{fig:entire}
\end{figure}

In fine-tuning BART for packet prediction (Figure \ref{fig:BARTarchi_a}), input packets are tokenized into token IDs, which are then embedded into high-dimensional vectors, capturing semantic information. These embedded tokens pass through both encoder and decoder layers of the BART model, employing multi-headed self-attention mechanisms followed by position-wise feedforward layers, effectively processing packet data as follows: 
\begin{equation}\begin{gathered} h_{0} = T W_{e} + W_{p}\\ h_{l} = tf\_block(h_{l-1}) \forall l\in [1,L]\end{gathered}\end{equation}

Where $T$ is a matrix of one-hot row vectors of the token indices in the sentence, $W_e$ is the token embedding matrix, and $W_p$ is the position embedding matrix, $L$ is the number of transformer blocks, and $h_l$ is the state at layer $l$. The encoder layers focus on capturing intricate contextual details and token relationships within the input sequence. The autoregressive decoder layers generate the output sequence token by token, considering previously generated tokens at each step. Additionally, each decoder layer conducts cross-attention over the final hidden layer of the encoder. This integration of encoder and decoder components enables BART to enhance its comprehension of the input packet iteratively. It generates numerical output, which is subsequently decoded into network packet format, producing coherent next network packets based on a comprehensive understanding of the preceding input tokens. More precisely, for an input network packet with N tokens, represented as $P$ = \{$p_1$ , $p_2$ , . . . , $p_N$\}.
This process computes the probability of token $t_k$, denoted as $p_n$, based on the preceding k-1 tokens:
\begin{equation} P_{N}(t_{k}| t_{1},...,t_{k-1})=softmax(W\upsilon h_{k-1}) \end{equation}
Where $h_k-1$ represents the encoded representation generated by the Transformer with the input of the previous tokens \{$t_1$,...,$t_k-1$\} and W$\upsilon$ signifies the trainable parameters.\par 

\subsubsection{LLM-based  packets classification} 
After generating next packets in the early stage of our intrusion prediction framework, we advance to fine-tune the BART model for packet classification. This involves using a labeled dataset containing normal and malicious network packets, as depicted in the development phase in Figure \ref{fig:archi3}. In the architecture shown in Figure \ref{fig:BARTarchi_b}, the fine-tuned BART model for supervised packet classification employs the same network packet input for both the encoder and decoder, using the final output representation. The objective is to maximize the likelihood of correctly classifying the class label $y$ given the packet features
$P$\{$f_1$ , $f_2$ , . . . , $f_N$\}, with the BART model's parameters denoted by $\theta$:
\begin{equation}
\mathcal L_{BARTencoder}(\theta) = -\frac{1}{N} \sum_{i=1}^{N} \log p(y| x_i ; \theta)
\end{equation}

During tokenization and embedding, BART utilizes various noise masking techniques and introduce controlled noise into the input sequence, aiding the model in learning robust representations by inferring missing or altered tokens within the sequence context. This approach ensures a comprehensive understanding of both forward and backward temporal context in network activities, enabling the model to distinguish from anomalies patterns. Throughout the fine-tuning process, BART adjusts its weights via backpropagation to minimize the loss function, aligning predicted class probabilities with ground truth labels. This iterative refinement process facilitates accurate classification of network packets. \par

\subsubsection{LLM-based next packet prediction assessing}

To evaluate BART's predicted next packets during the initial phase, we fine-tune another LLM concurrently. Specifically, we fine-tune BERT for the packet-pair classification task to assess BART's output. BERT masks a portion of input tokens, replacing them with a "[MASK]" token, and then predicts the original identities of these masked tokens using contextual information. The objective function maximizes the log-likelihood of predicting the correct tokens within the masked positions as follows:
\begin{equation}
\mathcal L_{BERT}(\theta)= \sum_{i=1}^N\operatorname{log}P{(xi\vert x_{masked};\operatorname\theta)} 
\end{equation}
Where $N$ is the number of masked positions, $x_i$ refers to the actual identity of the masked token at position i, $x_{masked}$ signifies the context surrounding the masked tokens, and $\theta$ represents BERT’s parameters.\par

BERT is chosen to evaluate BART's generated next packets due to its bidirectional context representation. Fine-tuning BERT involves preparing a labeled dataset, tokenizing, embedding, and adding special tokens for classification. The Masked Language Model predicts masked tokens, and a classification layer is added for packet-pair classification. During BART evaluation, the fine-tuned BERT model considers current and next packets, outputting probabilities for successive or non-successive classes.
 \par

\section{Performance Evaluation}\label{SIM}

In this section, we introduce the dataset employed, elaborate on the fine-tuning experiments and configurations, and subsequently showcase the results obtained. We then engage in a discussion concerning the performance metrics associated with our intrusion prediction framework.
\subsection{Dataset}

We utilized the CICIoT2023 dataset \cite{s23135941}, a realistic IoT attack dataset, to fine-tune BART and BERT, for next packet prediction, assessment, and classification tasks. This dataset, created in a controlled lab environment with 105 devices, includes 33 different attacks categorized into DDoS, DoS, Recon, Web-based, brute force, spoofing, and Mirai. We selected five attack types based on their real-world prevalence. Extracting packet features from raw network traffic using Tranalyzer \cite{tranalyzer}, we focused on layer 2, 3, and 4 protocol fields of the TCP/IP stack, resulting in 71 features, of which 26 were carefully chosen after feature selection techniques. To ensure feature quality, constant and quasi-constant features were removed using a minimum Variance Threshold of 25\%, and highly correlated features $(> 98\%)$ were eliminated through a Pearson correlation filter. This process ensured a refined set of relevant features for analysis, with each task requiring specific data preparation methods.

\subsection{Experiments} 

In our experiment, we utilized the CICIoT2023 dataset for three primary tasks: predicting the next packet, evaluating this prediction, and classifying packets. Leveraging
the HuggingFace transformer Python libraries \cite{wolf-etal-2020-transformers}, we fine-tuned BART for next packet prediction and BERT for prediction assessment in the Google Colab cloud environment. For packet classification, we fine-tuned BART on the Kaggle computational notebook platform using Python 3 and available GPU computing resources. \par

\subsubsection{Fine-tuning BART for next packets prediction} 

We utilized unlabeled network packets as a training dataset in fine-tuning BART for the next packet prediction task. Each dataset row includes the current packet and its next packet from the same network flow, aiding the model in understanding flow boundaries. The goal of this fine-tuning process is to improve the model's capacity to interpret and recognize the context and patterns of network traffic, ensuring accurate prediction of next network packets. BART employs byte pair encoding (BPE) for tokenizing network packets. We adjust the BART prediction function to generate a fixed-length representation of the network packet, maintaining consistency in the number of extracted features for each input packet. This adjustment aims to produce coherent, contextually relevant output while minimizing randomness. We choose to fine-tune the base version of the BART model, comprising 6 layers in both the encoder and decoder, with a total of 139 million parameters.\par

\subsubsection{Fine-tuning BERT for Packet-Pair Classification}

We performed supervised binary classification to evaluate BART’s predicted network packets by determining the sequence of two network packets. We created a dataset of packet pairs, categorized as successive or non-successive, derived from the CICIoT2023 dataset, which includes both normal and attack traffic. We fine-tuned the distilbert-base-uncased variant of the BERT model with 6 layers, 768 dimensions, 12 heads, and 66 million parameters for this pair-packet classification task. The model, trained for 15 epochs with a learning rate of 5e-5 and a batch size of 128 using the Adam optimizer, was optimized to classify packet pairs as successive or non-successive and evaluated alongside BART-predicted packets. An early stopping mechanism was applied to prevent overfitting.\par

\subsubsection{Fine-tuning BART for packets classification} 
Using two NVIDIA Tesla T4 GPUs with 15GB of memory each, we fine-tuned the base BART model for binary classification on labeled network packets from the CICIoT2023 dataset. The data was categorized into benign and specific attack types, with the distribution of normal and attack samples detailed in Table \ref{tab:distSamp}. 

\begin{table}[!ht]
\small
\centering
\caption{Dataset samples distribution}
\begin{tabular}{@{}ll@{}}
\toprule
Traffic Type & Nb. Samples\\ \midrule
Normal         & 1079391   \\
DDoS       & 55462   \\
Browser Hijacking       & 43414   \\
Command Injection       & 36323   \\
XSS       & 22838   \\
Backdoor Malware       & 19411   \\ \bottomrule
\end{tabular}
\label{tab:distSamp}
\end{table}

Subsequently, we split the data into training and validation subsets and fine-tuned BART for next packet classification over 4 epochs with a batch size of 2, using the Adam optimizer and 31,243 network packet instances on a Google Colab NVIDIA Tesla T4 GPU with 12 GB memory. The model parameters were optimized to minimize categorical cross-entropy loss. After training, we evaluated the model on the test set, assessing accuracy, precision, recall, and F1-score, and used the trained model to classify network packets.
\subsection{Results}
To assess the performance and predictive capabilities of our intrusion prediction framework, we examined various metrics including accuracy, recall, precision, and F1-score. We assumed that the labeled data from the CICIoT2023 dataset accurately represents real-world network traffic patterns.\par

The fine-tuned BERT model achieved 92.80\% accuracy, high precision and recall, and an F1 score of 96.10\%, as shown in Table \ref{tab:PP_perf}. This demonstrates the robust performance of the fine-tuned BERT model in discerning network packet succession. It identified 73\% of BART’s predicted packets as valid next network packets. Although these performances are promising, our framework needs further refinement. The prediction and assessment effectiveness could be improved by expanding the dataset with more diverse attack scenarios, conducting extensive validation across different network environments, and optimizing the models to enhance prediction accuracy and generalization capabilities.


\begin{table}[!ht]
\centering
\caption{Packet-Pair Classification Task results}
\begin{tabular}{@{}llll@{}}
\toprule
Accuracy & Precision & Recall & F1-Score\\ \midrule
92.80\%  & 94.46\% & 97.79\% & 96.10\%   \\
\bottomrule
\end{tabular}
\label{tab:PP_perf}
\end{table}

We conducted a binary classification to distinguish between malicious and normal packets. The confusion matrix obtained is depicted in Figure \ref{fig:bart_roc} . The Receiver Operating Characteristic (ROC) curve in Figure 4 shows the model’s ability to differentiate between positive and negative rates with 98\% separation, as indicated by the Area Under the Curve (AUC) value. The BART binary classification model achieved a testing accuracy of 98\% as presented in Table \ref{tab:lstm_report}, demonstrating exceptional precision, recall, and F1-score for both benign and attack traffic. Furthermore, leveraging BART’s generation capabilities to recognize and adapt to new patterns, our framework effectively generalizes to novel attacks, accurately identifying previously unseen threats.

\begin{figure}[ht!]
    \centering
    \includegraphics[scale=0.4]{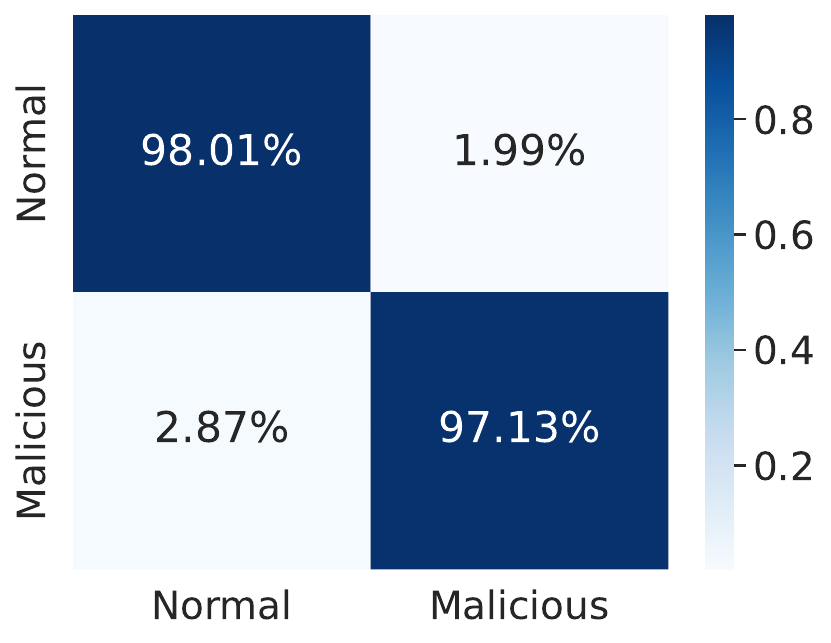}
    \caption{Confusion matrix of BART's binary classification}
    \label{fig:bart_cm}
\end{figure}

\begin{figure}[ht!]
    \centering
    \includegraphics[scale=0.4]{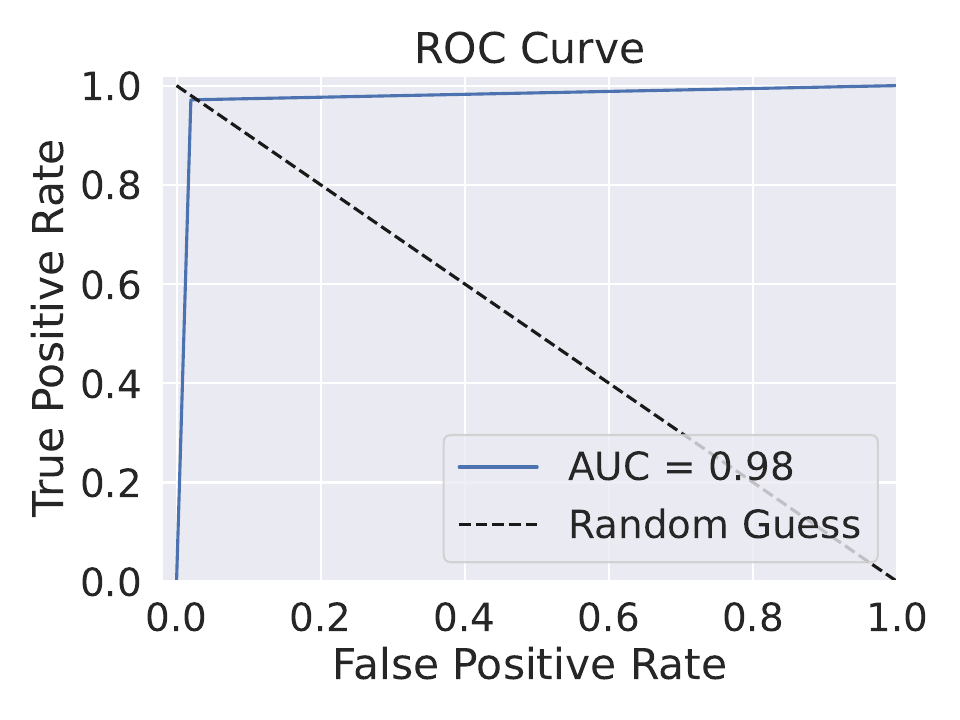}
    \caption{BART’s intrusion detection ROC}
    \label{fig:bart_roc}
\end{figure}


\begin{table}[!ht]
\centering
\caption{BART's packets binary classification report}
\begin{tabular}{@{}llll@{}}
\toprule
& Precision & Recall & F1-score \\ \midrule
Normal  & 0.9786 & \textbf{0.9868} & \textbf{0.9827}   \\
Malicious  & \textbf{0.9867} & 0.9784 & 0.9825   \\
\midrule
Accuracy & \multicolumn{3}{c}{\textbf{0.9826}} \\
\bottomrule
\end{tabular}
\label{tab:lstm_report}
\end{table}

\section{Conclusion} \label{CON}

IoT connectivity surge reveals reliance on IDSs is insufficient, as damage often precedes effective mitigation. Our paper presents a framework that enhances IoT security by accurately predicting and classifying network packets through fine-tuning BART and BERT LLMs, enabling effective intrusion prediction. This forward-looking approach enables prompt implementation of mitigation measures prior to attack occurrences. Our experiments prove the framework's robustness, achieving 98\% accuracy despite limited data during fine-tuning. Future work will assess the framework with diverse IoT datasets to ensure effectiveness against modern cybersecurity challenges.

\section*{Acknowledgment}

This work was supported by the 5G-INSIGHT bilateral project (ID: 14891397) / (ANR-20-CE25-0015-16), funded by the Luxembourg National Research Fund (FNR), and by the French National Research Agency (ANR).


\bibliographystyle{unsrt}
\bibliography{ref}

\end{document}